\documentclass[12pt,a4paper]{article}
\usepackage{amsmath,amsfonts,amsthm}
\allowdisplaybreaks[3]
\numberwithin{equation}{section}
\newtheorem{theorem}{Theorem}[section]

\newtheorem{definition}[theorem]{Definition}
\usepackage[body={16cm,23cm}]{geometry}
\usepackage[usenames,dvipsnames]{color}

\newcommand{\Lf}{{\mathbf L}}
\newcommand{\Lfb}{\overline{\mathbf L}}
\newcommand{\Kf}{{\mathbf K}}

\newcommand{\Ts}{{\mathsf T}}

\newcommand{\Pst}{\tilde{\mathsf P}}

\newcommand{\Ws}{{\mathsf W}}

\begin{document}
\title{Generic triangular solutions of the reflection equation: $U_{q}(\widehat{sl_2})$ case
}
\author{Zengo Tsuboi
\\[8pt]
{\sl
Laboratory of physics of living matter, 
}
 \\
 {\sl
School of biomedicine, 
Far Eastern Federal University, 
}
\\
{\sl
Sukhanova 8, Vladivostok  690950, Russia
}
\\[8pt]
{\sl
Osaka city university advanced mathematical institute,
}
\\
{\sl
3-3-138 Sugimoto, Sumiyoshi-ku Osaka 558-8585, Japan
}
} 
\date{}
\maketitle
\begin{abstract}
We consider intertwining relations of the triangular $q$-Onsager algebra, 
and obtain  generic triangular boundary $K$-operators in terms of the Borel subalgebras of 
$U_{q}(sl_2)$. These $K$-operators solve the  reflection equation. 
\end{abstract}

\vspace{10pt}

\noindent Journal of Physics A: Mathematical and Theoretical 53 (2020) 225202
\\
\noindent https://doi.org/10.1088/1751-8121/ab8853
\section{Introduction}
The reflection equation \cite{Ch84} is a fundamental object in 
quantum integrable systems with open boundary conditions \cite{Skly}. It 
 has the following expression: 
\begin{align}
R_{12} \left(\frac{y}{x}\right) K_{1}(x) \overline{R}_{12} \left( xy \right) 
 K_{2}(y) 
=K_{2}(y) 
 R_{12} \left(\frac{1}{xy} \right)  K_{1}(x) \overline{R}_{12} \left(\frac{x}{y}\right), 
\quad  x,y \in \mathbb{C}^{\times},
\label{refeq00}
\end{align}
where $R(x)$ and $\overline{R}(x)$ are solutions (R-matrices) of the Yang-Baxter equation and $K(x)$ is 
 a K-matrix. The indices 1 and 2 denote the space on which the operators act non-trivially. 
 In particular, 
there is a $2 \times 2$  matrix solution \cite{DeV,GZ,MN} of the reflection equation 
associated with the $4 \times 4$ R-matrices of the 6-vertex model
(for the $2$-dimensional fundamental representation  of  $U_{q}(\widehat{sl_{2}})$: 
see \eqref{Rmat1} and \eqref{Rmat2}):
\begin{align}
K (x) & =
\begin{pmatrix}
x^{s_{0}} \epsilon_{+}  +  x^{-s_{1}} \epsilon_{-} & 
\frac{k_{+}(x^{s}-x^{-s})}{q-q^{-1}}  \\
\frac{k_{-}(x^{s}-x^{-s})}{q-q^{-1}}  & 
x^{-s_{0}}\epsilon_{+}  +  x^{s_{1}} \epsilon_{-}
 \end{pmatrix} ,
 \label{Kmat-sc0}
\end{align}
where  $k_{\pm}$ and $\epsilon_{\pm}$ are scalar parameters.
These R-matrices are evaluations of more general operators 
called $L$-operators:  
$\Lf_{12}(x),\overline{\Lf}_{12}(x)  \in U_{q}(sl_{2}) \otimes \mathrm{End}({\mathbb C}^{2})$ 
(see \eqref{Lgen1} and \eqref{Lgen2}). 
Namely, they are given by 
$R_{12}(x)=(\pi \otimes 1)\Lf_{12}(x)$, $\overline{R}_{12}(x)=(\pi \otimes 1)\Lfb_{12}(x)$, 
 where $\pi$ is the fundamental representation of $U_{q}(sl_{2})$.
In this context, 
a natural problem is to explore the solutions of the reflection equation associated with 
the L-operators: 
\begin{align}
\Lf_{12} \left(\frac{y}{x}\right) \Kf_{1}(x) \overline{\Lf}_{12} \left( xy \right) 
 K_{2}(y) 
=K_{2}(y) 
 \Lf_{12} \left(\frac{1}{xy} \right)  \Kf_{1}(x) \overline{\Lf}_{12} \left(\frac{x}{y}\right),
\label{refeq20}
\end{align}
where  $\Kf(x)$ is a generic
\footnote{We use the term `generic' since the K-operator is written in terms of generators of 
the symmetry algebra and is independent of the representations. 
In addition, the reason why we use the term `generic' instead of a more familiar term `universal' 
is that we are considering the problem under the evaluation map from
 $U_{q}(\widehat{sl_{2}})$ to $U_{q}(sl_{2})$.}
 K-operator  in $U_{q}(sl_{2})$. 
In this paper, we propose generic triangular solutions $\Kf(x)$ of \eqref{refeq20} associated with 
the triangular K-matrices (\eqref{Kmat-sc0} for $k_{+}=0$ or $k_{-}=0$ ) 
in terms of the elements of the Borel subalgebras of  $U_{q}(sl_{2})$ (see \eqref{Kgen-up}, \eqref{Kgen-lo}, \eqref{Kgen-upd} and \eqref{Kgen-lou}). 
Evaluation of the generic K-operator in  
the fundamental representation of $U_{q}(sl_{2})$ reproduces the $2 \times 2$ 
triangular K-matrices \eqref{Kmat-sc0}: $K(x)=\pi(\Kf(x))$. 
In the context of Baxter Q-operators for integrable systems with open boundaries, 
  generic diagonal K-operators
 \footnote{These solutions were generalized to (a quotient of) higher rank algebra case \cite{Ts18}.}
  ($k_{+}=k_{-}=0$ case) for $U_{q}(\widehat{sl_{2}})$ 
 were previously proposed in \cite{BT18}. 
This paper extends these to the triangular case  in part. 
Although it is beyond the scope of the present paper, we expect that 
our results will be useful to construct Baxter Q-operators for integrable systems with triangular boundaries 
(by taking limits of K-operators as discussed in \cite{BT18}). 
We also remark that there are solutions of the reflection equation for the symmetric tensor 
 representations of $U_{q}(A^{(1)}_{n-1})$ \cite{KOY18} and $U_{q}(\widehat{sl_{2}})$ \cite{ML19}. 
 The directions of their results \cite{KOY18,ML19} are different from ours in that their solutions are not expressed 
 in terms of generators of symmetry algebras and thus depend on representations. 
As for the rational case ($q=1$ case; the XXX-model associated with the Yangian $Y(sl_2)$),  
diagonal K-operators for Baxter Q-operators appeared first in \cite{FS15},
 and a generic non-diagonal K-operator
  was proposed recently
\footnote{In 2016, we were informed by 
S.\ Belliard that he obtained a generic non-diagonal K-operator
for the rational case.}
 in \cite{FGK19}. 

In section 2, we review the quantum algebras $U_{q}(\widehat{sl_{2}})$ and $U_{q}(sl_{2})$ and 
associated R-and L-operators. In section 3, we recall the triangular q-Onsager algebra $ O^{t}_{q}(\widehat{sl_{2}})$ \cite{BB17}, which 
is the underlying symmetry for the XXZ-spin chain with triangular boundary conditions. 
It is a co-ideal subalgebra (cf.\ \cite{MRS}) of $U_{q}(\widehat{sl_{2}})$. 
 We solve the intertwining relations of the form
\begin{align}
\mathsf{ev}_{x^{-1}}(a)\Kf(x) &= 
 \Kf(x)\mathsf{ev}_{x}(a) 
 \quad \text{for} \quad a \in  O^{t}_{q}(\widehat{sl_{2}}), 
 \label{int-0}
\end{align}
where $\mathsf{ev}_{x}$ is an evaluation map $\mathsf{ev}_{x}: O^{t}_{q}(\widehat{sl_{2}}) \to U_{q}(sl_{2})$
with the spectral parameter $x \in {\mathbb C}^{\times}$. We show
\footnote{A standard procedure to connect intertwining relations and the reflection equation 
 will be to consider irreducible 
representations and use Schur's lemma (cf. \cite{DM03}). Here we consider the problem on the level of the algebra.} 
that  solutions of these intertwining relations also solve 
   the reflection equation \eqref{refeq20}. In section 4, we discuss the connection to 
  the q-Onsager algebra \cite{Te03,Ba04}, which is the symmetry algebra for the case $k_{+}k_{-} \ne 0$. 
  In the appendix, we review miscellaneous formulas which follow from the q-deformed Hadamard formula \cite{KT91}.  

Throughout this paper, we assume that the deformation parameter is of the form $q$  is not a root of unity. 
We also identify 
 the multiplicative unit element ${\mathbf 1}$ of the quantum algebra multiplied by a complex number $b\in {\mathbb C}$ with $b$ 
 by an algebra embedding from the field of scalars into the associative algebra: $b{\mathbf 1}=b$. 
We use the following notation:

\vspace{6pt}

\noindent 
{\bf Notation:} 
\begin{itemize}
\item
For any elements  $X,Y$ of the quantum algebras, we define the q-commutator by   
$[X,Y]_{q}=XY-qYX$. In particular, we set $[X,Y]_{1}=[X,Y]$.

 %
 \item 
 We introduce an expression  
$(x;q)_{k}=\prod_{j=0}^{k-1}(1-xq^{j})$. In particular, we define 
$(x;q)_{\infty}=\lim_{k \to \infty}(x;q)_{k}=\prod_{j=0}^{\infty}(1-xq^{j})$ for $|q|<1$. 
For more detail, see for example, page 38 in \cite{KS97}. 

\item
We define a 
q-analogue of the exponential function by   
$\exp_{q}(x)=1+\sum_{k=1}^{\infty} \frac{x^{k}}{(k)_{q} !}=((1-q)x;q)^{-1}_{\infty}$ 
for $|q|<1$, $|(1-q)x|<1$, where 
$(k)_{q}! =(1)_{q}(2)_{q}\cdots (k)_{q}$, 
$ (0)_{q}!=1$, $(k)_{q}= (1-q^{k})/(1-q)$. 
For more detail, see for example, page 47 in \cite{KS97}. 

\item 
We will use free parameters
  $s_{0},s_{1} \in {\mathbb Z}$. In particular, we set $s=s_{0}+s_{1}$. 
\end{itemize}
 \section{Quantum algebras and L-operators}
In this section, we review quantum algebras and associated R-and L-operators. 
We basically follow the convention in \cite{BT18}. 
We also refer to  \cite{CP95,KS97} for review on this subject.
\subsection{The quantum affine algebra $U_{q}(\widehat{sl_2})$}
Let us start from the definition of the algebra. 
\begin{definition}
The quantum affine algebra $U_{q}(\widehat{sl_2})$ (at level 0, i.e. the quantum loop algebra)  is a Hopf algebra
generated by the generators
 $e_{i},f_{i},q^{\xi h_{i}}$ for 
$i \in \{0,1 \}$ and $\xi \in {\mathbb C}$ obeying the following relations:
\begin{align}
& q^{0 h_{i}}=q^{0}=1, \quad q^{\xi h_{i}}q^{\eta h_{i}}=q^{(\xi +\eta )h_{i}},  
\quad q^{\xi h_{0}}q^{\xi h_{1}}=1,
\label{sl2h-def1}
\\[6pt]
&[e_{i},f_{j}]=\delta_{ij} \frac{q^{h_{i}} -q^{-h_{i}} }{q-q^{-1}},
\quad q^{\xi h_{i}} e_{j}q^{-\xi h_{i}} =q^{\xi a_{ij} }e_{j}, \quad
q^{\xi h_{i}} f_{j}q^{-\xi h_{i}}  =q^{-\xi a_{ij} }f_{j}, 
\label{sl2h-def2}
\\[6pt]
&[e_{i},[e_{i},[e_{i},e_{j}]_{q^{2}}]]_{q^{-2}}=
[f_{i},[f_{i},[f_{i},f_{j}]_{q^{-2}}]]_{q^{2}}=0
\qquad i \ne j , 
\quad \xi, \eta \in {\mathbb C},
\label{sl2h-def3}
\end{align}
where $(a_{ij})_{0 \le i,j\le 1}$ is the
Cartan matrix
\begin{align}\nonumber
(a_{ij})_{0 \le i,j\le 1}=
\begin{pmatrix}
2& -2 \\
-2 & 2
\end{pmatrix}
.
\end{align}
\end{definition}
%
The algebra has automorphisms $\sigma$ and $\tau $  defined by 
\begin{align}
\begin{split}
& \sigma (e_{0})=e_{1}, \qquad \sigma (f_{0})=f_{1}, \qquad \sigma (q^{\xi h_{0}})=q^{\xi h_{1}},  
\\[6pt]
& \sigma (e_{1})=e_{0}, \qquad \sigma (f_{1})=f_{0}, \qquad \sigma (q^{\xi h_{1}})=q^{\xi h_{0}}, 
\qquad \sigma(q)=q 
\end{split}
\label{auto1}
\end{align}
and 
\begin{align}
& \tau (e_{i})=f_{i}, \quad \tau (f_{i})=e_{i}, \quad \tau (q^{\xi h_{i}})=q^{-\xi h_{i}},   
\quad \tau(q)=q,
\quad i=0,1.
\label{auto2}
\end{align}
The algebra also has an  anti-automorphism $\iota$  defined by
\footnote{For any $a\in {\mathbb C}$ and a Cartan element ${\mathcal H}$, 
we denote $q^{a}q^{{\mathcal H}}$ as $q^{a+{\mathcal H}}$. } 
\begin{align}
& \iota(e_{i})=q^{-1-h_{i}}f_{i}, \quad \iota(f_{i})=e_{i}q^{1+h_{i}}, \quad \iota(q^{\xi h_{i}})=q^{\xi h_{i}},
\quad \iota(q)=q,
\quad i=0,1.
\label{a-auto1}
\end{align}
Recall that this means $\sigma(ab)=\sigma(a)\sigma(b)$, $\tau(ab)=\tau(a)\tau(b)$ and 
 $\iota(ab)=\iota(b)\iota(a)$ 
for $a,b \in U_{q}(\widehat{sl_2})$. 
We use the following co-product
  $ \Delta : U_{q}(\widehat{sl_2}) \to U_{q}(\widehat{sl_2}) \otimes U_{q}(\widehat{sl_2})$:
\begin{align}
\Delta (e_{i})&=e_{i} \otimes 1 + q^{-h_{i}} \otimes e_{i}, \nonumber\\
\Delta (f_{i})&=f_{i} \otimes q^{h_{i}} + 1 \otimes f_{i},\label{copro-h} \\
\Delta (q^{\xi h_{i}})&=q^{\xi h_{i}} \otimes q^{\xi h_{i}}. \nonumber
\end{align}
We will also use the opposite co-product defined by
\begin{align}
\Delta'={\mathfrak p}\circ \Delta,\qquad {\mathfrak p}\circ 
(X\otimes Y)=
Y\otimes X,\qquad X,Y\in U_{q}(\widehat{sl_{2}}).
\end{align}
Anti-pode, co-unit and grading element $d$ are not  used  in this
paper.

The Borel subalgebra ${\mathcal B}_{+}$  and ${\mathcal B}_{-}$ are generated by
$e_{i}, q^{\xi h_{i}} $ and $f_{i},  q^{\xi h_{i}}$, respectively, where
$i \in \{0,1 \}$, $\xi \in {\mathbb C}$.
There exists a unique element \cite{Dr85,KT92} 
${\mathcal R}$ in a completion of $ {\mathcal B}_{+} \otimes {\mathcal B}_{-} $ 
called the universal R-matrix which satisfies the following 
relations
\begin{align}
\begin{split}
\Delta'(a)\ {\mathcal R}&={\mathcal R}\ \Delta(a)
\qquad \text{for} \quad \forall\ a\in U_{q}(\widehat{{sl}_{2}}),   \\[6pt]
(\Delta\otimes 1)\, 
{\mathcal R}&={\mathcal R}_{13}\, {\mathcal R}_{23}, \\[6pt]
(1\otimes \Delta)\, {\mathcal R}&={\mathcal R}_{13}
{\mathcal R}_{12} ,
\end{split}
\label{R-def}
\end{align}
where
\footnote{We will use similar notation for the L-operators 
to indicate the space on which they non-trivially act.} 
${\cal R}_{12}={\cal R}\otimes 1$, ${\cal R}_{23}=1\otimes {\cal R}$,
${\cal R}_{13}=({\mathfrak p}\otimes 1)\, {\cal R}_{23}$.
The Yang-Baxter equation 
\begin{align}
{\mathcal R}_{12}{\mathcal R}_{13}{\cal R}_{23}=
{\mathcal R}_{23}{\mathcal R}_{13}{\mathcal R}_{12} \label{YBE}
\end{align}
follows from these relations \eqref{R-def}. 

\subsection{The quantum algebra $U_{q}(sl_2)$}
\begin{definition}
The quantum algebra $U_{q}(sl_2)$ is  generated by the elements $E, F, q^{\xi H}$ 
for $\xi \in {\mathbb C}$ obeying the following relations:
\begin{align}
&
q^{0H}=q^{0}=1, \quad 
q^{\xi H}q^{\eta H}=q^{(\xi+\eta )H}, \quad 
q^{\xi H}Eq^{-\xi H}=q^{2\xi}E , \qquad q^{\xi H}Fq^{-\xi H}=q^{-2\xi}F ,
\nonumber
\\ &
[E, F] = \frac{q^{H} - q^{-H} }{q-q^{-1}}, 
\quad \xi,\eta  \in {\mathbb C}.
\label{HEF-sl2}
\end{align}
\end{definition}
The upper (resp.\ lower) Borel subalgebra is generated by $E,q^{\xi H}$ (resp.\ $F,q^{\xi H}$). 
The Casimir element 
\begin{align}
C=FE+\frac{q^{H+1} + q^{-H-1}}{(q-q^{-1})^{2}} 
=EF+\frac{q^{H-1} + q^{-H+1}}{(q-q^{-1})^{2}}
\label{Casimir}
\end{align}
is central in $U_{q}(sl_2)$. 
There are an automorphism 
\begin{align}
\sigma(E)= F,
\qquad 
\sigma(F)= E,
\qquad 
\sigma(q^{\xi H})= q^{-\xi H}
\qquad 
\sigma(q)=q,
\end{align}
 and an anti-automorphism 
\begin{align}
& 
\iota(E)= q^{-H-1}F,
\qquad 
\iota(F)=Eq^{H+1},
\qquad 
\iota(q^{\xi H})= q^{\xi H}, 
\qquad
\iota(q)=q
 \label{t-sl2}
\end{align}
of the algebra.
They are $U_{q}(sl_2)$ analogues  of \eqref{auto1} and \eqref{a-auto1}, respectively. 
There is an algebra homomorphism called evaluation map
\footnote{We emulate \cite{BGKNR10} and consider 
the general gradation of the algebra.} $\mathsf{ev}_{x}$:
$U_{q}(\widehat{sl_2}) \mapsto U_{q}(sl_2)$,
\begin{align}
\begin{split}
& e_{0} \mapsto x^{s_{0}}F,  \qquad
 f_{0} \mapsto x^{-s_{0}} E, \qquad
q^{\xi h_{0}}  \mapsto q^{-\xi H},
\\
& e_{1} \mapsto x^{s_{1}} E,  \qquad
f_{1} \mapsto x^{-s_{1}} F, \qquad
q^{\xi h_{1}}  \mapsto q^{\xi H},
\end{split}
\label{eva}
\end{align}
where $x  \in {\mathbb C}^{\times}$ is the spectral parameter. 
We set
\footnote{The operations to permute the free integer parameters $(s_{0},s_{1})$ in the evaluation map are originally independent of the 
automorphisms or anti-automorphisms of the algebras. But we synchronize these and purposely use the same notation. 
A similar remark holds true for \eqref{para1} and \eqref{para2}.}
\begin{align}
&  \sigma (s_{0})=s_{1}, \quad \sigma (s_{1})=s_{0}, \label{sig-con} 
\\[6pt]
&\iota(s_{0})=s_{0}, \quad \iota(s_{1})=s_{1}. 
\label{iot-con}
\end{align}  
One can check consistency of these: 
$ \sigma\circ \mathsf{ev}_{x}=\mathsf{ev}_{x} \circ \sigma $ and 
$ \iota\circ \mathsf{ev}_{x}=\mathsf{ev}_{x^{-1}} \circ \iota $. 
The fundamental representation $\pi$ of $U_{q}(sl_2)$ is given by 
$\pi(E)=E_{12}$, $\pi(F)=E_{21}$ and $\pi(q^{\xi H})=q^{\xi}E_{11}+q^{-\xi}E_{22}$ , where
$E_{ij}$ is the $2 \times 2$ matrix unit whose
$(k,l)$-element is $\delta_{i,k}\delta_{j,l}$. 
The composition 
$\pi_{x}=\pi \circ \mathsf{ev}_{x}$
gives an evaluation representation of $U_{q}(\widehat{sl_2})$.
In case we consider the fundamental representation,  we define 
 an algebra automorphism $\sigma$ and an algebra anti-automorphism $\iota$
 of the algebra of $2\times 2$ matrices over ${\mathbb C}$ by 
\begin{align}
\sigma(E_{ij})&=E_{3-i,3-j}, 
\\[6pt]
\iota(E_{ij})&=E_{ji}, \qquad i,j=1,2.
\end{align}
We have an identity of algebra homomorphisms $\pi \circ \sigma = \sigma \circ \pi$ and an identity of algebra
 anti-homomorphisms $\pi \circ \iota = \iota \circ \pi,$ which justifies our use of the same symbol for different maps.
\subsection{L-operators}
The so-called L-operators are images of the universal R-matrix, which are given by 
$\Lf (xy^{-1})=\phi(xy^{-1})(\mathsf{ev}_{x} \otimes \pi_{y}) {\mathcal R}$, 
$\overline{\Lf}  (xy^{-1})=\phi(x^{-1}y)(\mathsf{ev}_{x} \otimes \pi_{y}) {\mathcal R}_{21}$, 
where $x,y \in {\mathbb C}^{\times}$, and $\phi(xy^{-1})$ is an overall factor whose explicit expression 
will not be used in this paper.  
They are solutions of the intertwining relations, which 
follow from \eqref{R-def}:
\begin{align}
& \left((\mathsf{ev}_{x} \otimes \pi_{y})  \Delta^{\prime }(a)\right) \Lf (xy^{-1}) =\Lf (xy^{-1}) 
\left((\mathsf{ev}_{x} \otimes \pi_{y})  \Delta(a)\right)  ,
 \label{intert1}
\\[6pt]
& \left((\mathsf{ev}_{x} \otimes \pi_{y})  \Delta(a)\right) \overline{\Lf} (xy^{-1}) =\overline{\Lf} (xy^{-1}) 
\left((\mathsf{ev}_{x} \otimes \pi_{y})  \Delta^{\prime }(a)\right)  
\qquad \forall a \in U_{q}(\widehat{sl_{2}}).
 \label{intert2}
\end{align}
Explicitly, they read
\begin{align}
\Lf (x)& =
\begin{pmatrix}
q^{\frac{H}{2}} -q^{-1} x^{s} q^{-\frac{H}{2}} & (q-q^{-1}) x^{s_{0}}F q^{-\frac{H}{2} } \\
(q-q^{-1}) x^{s_{1}} E q^{\frac{H}{2} } & q^{- \frac{H}{2}} -q^{-1} x^{s} q^{\frac{H}{2}}
 \end{pmatrix} ,
 \label{Lgen1}
 \\[8pt]
\overline{\Lf} (x)& =
\begin{pmatrix}
q^{\frac{H}{2}} -q^{-1} x^{-s} q^{-\frac{H}{2}} & (q-q^{-1}) x^{-s_{1}}F q^{-\frac{H}{2} } \\
(q-q^{-1}) x^{-s_{0}} E q^{\frac{H}{2} } & q^{- \frac{H}{2}} -q^{-1} x^{-s} q^{\frac{H}{2}}
 \end{pmatrix} .
 \label{Lgen2}
\end{align}
One can check that these L-operators satisfy the unitarity and the crossing unitarity conditions 
(cf.\ eqs.\ (3.5) and (3.6) in \cite{BT18}). 
Evaluating the first space of these L-operators in the fundamental representation, 
we obtain R-matrices of the 6-vertex model.
\begin{align}
R(x)=
q^{\frac{1}{2}}(\pi \otimes 1)\Lf (x)& =
\begin{pmatrix}
q-q^{-1}x^{s} & 0 & 0 & 0 \\
0 & 1-x^{s} &  (q-q^{-1}) x^{s_{1}} & 0 \\
0 & (q-q^{-1}) x^{s_{0}} & 1-x^{s}  & 0 \\
 0 & 0 & 0 & q-q^{-1}x^{s}
 \end{pmatrix} ,
 \label{Rmat1}
 \\[8pt]
\overline{R}(x)=
q^{\frac{1}{2}}(\pi \otimes 1)\overline{\Lf} (x)& =
\begin{pmatrix}
q-q^{-1}x^{-s} & 0 & 0 & 0 \\
0 & 1-x^{-s} &  (q-q^{-1}) x^{-s_{0}} & 0 \\
0 & (q-q^{-1}) x^{-s_{1}} & 1-x^{-s}  & 0 \\
 0 & 0 & 0 & q-q^{-1}x^{-s}
 \end{pmatrix} .
 \label{Rmat2}
\end{align}
These satisfy Yang-Baxter relations, which follow from \eqref{YBE}:
\begin{align}
& R_{12}(x/y)R_{13}(x/z)R_{23}(y/z)=R_{23}(y/z)R_{13}(x/z)R_{12}(x/y),
\quad 
x,y,z \in {\mathbb C}^{\times},
\label{YBE2}
\\[5pt]
& {\overline R}_{12}(x/y){\overline R}_{13}(x/z){\overline R}_{23}(y/z)={\overline R}_{23}(y/z){\overline R}_{13}(x/z){\overline R}_{12}(x/y),
\label{YBE3}
\\[6pt]
&\Lf_{12}(x/y)\Lf_{13}(x/z)R_{23}(y/z)=R_{23}(y/z)\Lf_{13}(x/z)\Lf_{12}(x/y),
\label{YBE4}
\\[6pt]
&\overline{\Lf}_{12} (x/y)\overline{\Lf}_{13}(x/z) \overline{R}_{23}(y/z)=\overline{R}_{23}(y/z)\overline{\Lf}_{13}(x/z)\overline{\Lf}_{12}(x/y) .
\label{YBE5}
\end{align}
One can check 
$(\sigma \otimes \sigma )\Lf(x)=\Lf(x)$,  $(\sigma \otimes \sigma )\overline{\Lf}(x)=\overline{\Lf}(x)$, 
$(\sigma \otimes \sigma )R(x)=R(x)$,  $(\sigma \otimes \sigma )\overline{R}(x)=\overline{R}(x)$, 
$(\iota \otimes \iota )\Lf(x)=\overline{\Lf}(x^{-1})$, $(\iota \otimes \iota )\overline{\Lf}(x)=\Lf(x^{-1})$, 
$(\iota \otimes \iota )R(x)=\overline{R}(x^{-1})$, $(\iota \otimes \iota )\overline{R}(x)=R(x^{-1})$. 
Thus \eqref{YBE2}-\eqref{YBE5} are invariant under the map $\sigma \otimes\sigma \otimes\sigma $; 
\eqref{YBE2} and \eqref{YBE3} swap and \eqref{YBE4} and \eqref{YBE5} swap  under the map $\iota \otimes \iota \otimes \iota $. 

\section{The reflection equation and its solutions}
\label{sec:RE}
In this section, we consider  intertwining relations of the triangular q-Onsager algebra
and obtain  generic K-operators in terms of the Borel subalgebras of  $U_{q}(sl_{2})$. 
These K-operators give solutions of  the reflection equation 
associated with the L-operators. 
\subsection{Reflection equation}
We start from the following form of the reflection equation \cite{Ch84} for the R-matrices 
\eqref{Rmat1} and \eqref{Rmat2}:
\begin{align}
R_{12} \left(\frac{y}{x}\right) K_{1}(x) \overline{R}_{12} \left( xy \right) 
 K_{2}(y) 
=K_{2}(y) 
 R_{12} \left(\frac{1}{xy} \right)  K_{1}(x) \overline{R}_{12} \left(\frac{x}{y}\right),
\label{refeq0}
\end{align}
where $x,y \in {\mathbb C}^{\times}$, $K_{1}(x)=K(x) \otimes 1$, $K_{2}(y)=1 \otimes K(y)$. 
The most general solution of the reflection equation \eqref{refeq0} is
 given  by (see \cite{DeV,GZ,MN})
\begin{align}
K (x) & =
\begin{pmatrix}
x^{s_{0}} \epsilon_{+}  +  x^{-s_{1}} \epsilon_{-} & 
\frac{k_{+}(x^{s}-x^{-s})}{q-q^{-1}}  \\
\frac{k_{-}(x^{s}-x^{-s})}{q-q^{-1}}  & 
x^{-s_{0}}\epsilon_{+}  +  x^{s_{1}} \epsilon_{-}
 \end{pmatrix} ,
 \label{Kmat-sc}
\end{align}
where  $k_{\pm}$ and $\epsilon_{\pm}$ are scalar parameters. We assume  $\epsilon_{+}\epsilon_{-} \ne 0$ 
since we will deal with solutions which contain $\epsilon_{+}^{-1}$ or $\epsilon_{-}^{-1}$. 
We would like to consider the 
reflection equation for the L-operators \eqref{Lgen1} and \eqref{Lgen2}: 
\begin{align}
\Lf_{12} \left(\frac{y}{x}\right) \Kf_{1}(x) \overline{\Lf}_{12} \left( xy \right) 
 K_{2}(y) 
=K_{2}(y) 
 \Lf_{12} \left(\frac{1}{xy} \right)  \Kf_{1}(x) \overline{\Lf}_{12} \left(\frac{x}{y}\right),
\label{refeq2}
\end{align}
and solve this with respect to the K-operator $\Kf(x)$. 
The reflection equation \eqref{refeq0} is the image of  \eqref{refeq2} 
for $\pi \otimes 1$.  
We set
\begin{align}
&\sigma(\epsilon_{+})=\epsilon_{-} ,
\qquad 
\sigma(\epsilon_{-})=\epsilon_{+} , 
\nonumber \\[6pt]
&\sigma(k_{+})=k_{-} \quad \text{for} \quad k_{+} \ne 0,
\qquad 
\sigma(k_{-})=k_{+} \quad \text{for} \quad k_{-} \ne 0;
\label{para1}
\\[6pt]
&\iota (\epsilon_{+})=\epsilon_{+}, \qquad 
\iota (\epsilon_{-})=\epsilon_{-}, \nonumber
\\[6pt]
&\iota(k_{+})=k_{-} \quad \text{for} \quad k_{+} \ne 0,
\qquad 
\iota(k_{-})=k_{+} \quad \text{for} \quad k_{-} \ne 0. 
\label{para2}
\end{align}
The reflection equations \eqref{refeq0} and \eqref{refeq2} 
are invariant under the action of $\sigma \otimes \sigma $ 
and $\iota \otimes \iota $  
if $k_{+}k_{-} \ne 0$; 
while those for the upper triangular K-matrix (\eqref{Kmat-sc} with $k_{-}=0$) and 
the lower triangular K-matrix (\eqref{Kmat-sc} with $k_{+}=0$)
swap one another. 
Thus we can derive the generic lower triangular K-operator from the upper triangular one by $\sigma $ or $\iota$. 
 \subsection{The triangular q-Onsager algebra}
 The triangular q-Onsager algebra is an underlying symmetry algebra of triangular solutions of the reflection equation. 
 \begin{definition}
  The triangular q-Onsager algebra $O^{t}_{q}(\widehat{sl_{2}})$ \cite{BB17} 
  is generated by the generators
\footnote{We assume the central element $\Gamma$ is $1$.}
 $\Ts_{0},\Ts_{1}, \Pst_{1}$ obeying the following relations
\begin{align}
& [\Ts_{1},[\Ts_{1},\Pst_{1}]_{q^{2}}]= k_{+} q (q+q^{-1})^{2} [\Ts_{0},\Ts_{1}], 
\quad 
[\Ts_{0},[\Ts_{0},\Pst_{1}]_{q^{-2}}]=k_{+} q^{-1} (q+q^{-1})^{2}[\Ts_{0},\Ts_{1}],
\nonumber 
\\[6pt]
& [\Ts_{1},\Ts_{0}]_{q^{-2}}=\epsilon_{+}\epsilon_{-}(1-q^{-2}),
\label{tri-alg}
\end{align}
where $k_{+},\epsilon_{+},\epsilon_{-} \in {\mathbb C}$. 
\end{definition}
The algebra $O^{t}_{q}(\widehat{sl_{2}})$ can be 
realized in terms of  the generators of $U_{q}(\widehat{sl_{2}})$ as follows. 
\footnote{The convention used in eq. (2.12)  in the original paper \cite{BB17} is related to \eqref{tri-real} 
by the automorphism  \eqref{auto2} of $U_{q}(\widehat{sl_{2}})$ and
the replacement $k_{\pm}\to k_{\mp}$ and $q \to q^{-1}$
(under the condition $h_{0}+h_{1}=c=0$, $\Gamma=1$.)}
  \begin{align}
  \Ts_0&= k_{+}qe_{1}q^{h_{1}} + \epsilon_{+} q^{h_1}
  \nonumber  \\[6pt]
\Ts_1&=  k_{+}f_0 + \epsilon_{-}q^{h_0}, 
\label{tri-real}
\\[6pt]
 \Pst_{1}&=-(q^{2}-q^{-2})(\epsilon_{-}qf_{1}q^{h_{0}}+
 \epsilon_{+}e_{0})+k_{+}q^{-1}([f_{1},f_{0}]_{q^{2}}+[e_{1},e_{0}]_{q^{2}})+\tilde{p} , \nonumber 
 \end{align}
 where $\tilde{p} \in {\mathbb C}$.
 %
%
From now on we identify $O^{t}_{q}(\widehat{sl_{2}})$ with the corresponding subalgebra of $U_{q}(\widehat{sl_{2}})$ 
and note that it is a right coideal, i.e.
$\Delta (O^{t}_{q}(\widehat{sl_{2}})) \subseteq O^{t}_{q}(\widehat{sl_{2}}) \otimes U_{q}(\widehat{sl_{2}})$
which follows from the following formulas for the generators:
 \begin{align}
 %
 \Delta(\Ts_{0})& =\Ts_{0} \otimes q^{h_{1}}+1 \otimes (k_{+}qe_{1}q^{h_{1}}), 
 \qquad 
 \Delta(\Ts_{1})=\Ts_{1} \otimes q^{h_{0}}+1 \otimes (k_{+}f_{0}), 
 \nonumber  \\[6pt]
 \Delta(\Pst_{1})&=\Pst_{1} \otimes 1 -(q^{2}-q^{-2})(\Ts_{1} \otimes f_{1}q^{1+h_{0}}+\Ts_{0} \otimes  e_{0})
 \nonumber \\ & \qquad \qquad 
 +1 \otimes k_{+}q^{-1}([f_{1},f_{0}]_{q^{2}}+[e_{1},e_{0}]_{q^{2}}).
 \end{align}
 Equivalently, the subalgebra $ O^{t}_{q}(\widehat{sl_{2}})$ is a left coideal with respect to the opposite co-product:  
 $ \Delta^{\prime}(O^{t}_{q}(\widehat{sl_{2}})) \subseteq  U_{q}(\widehat{sl_{2}}) \otimes O^{t}_{q}(\widehat{sl_{2}}) $. 
 One can apply the evaluation map to \eqref{tri-real} and obtain a homomorphism $O^{t}_{q}(\widehat{sl_{2}}) \mapsto U_{q}(sl_{2})$:
 \begin{align}
  \mathsf{ev}_{x}(\Ts_0)&= k_{+}qx^{s_1} Eq^{H}  + \epsilon_{+} q^{H} , 
  \nonumber 
  \\[6pt]
\mathsf{ev}_{x}(\Ts_1)&= k_{+}x^{-s_0}E + \epsilon_{-} q^{-H} ,
\label{realop2}
\\[6pt]
 \mathsf{ev}_{x}(\Pst_{1})&=-(q^{2}-q^{-2})F(\epsilon_{-}qx^{-s_{1}}q^{-H}+\
  \epsilon_{+}x^{s_{0}})
  \nonumber \\
  & \quad +
  k_{+}\left(-(q-q^{-1}) (x^{s}+x^{-s})C+\frac{q+q^{-1}}{q-q^{-1}}(x^{s}q^{H}+x^{-s}q^{-H}) \right)+\tilde{p}.
  \nonumber 
 \end{align}
 We remark that the algebras generated by  
\{ $\Ts_{0}^{\prime}=\sigma(\Ts_{0})$, $\Ts_{1}^{\prime}=\sigma(\Ts_{1})$, $\Pst_{1}^{\prime}=\sigma(\Pst_{1})$\} 
and  
  \{$\Ts_{1}^{\prime}=\iota(\Ts_{0})$, $\Ts_{0}^{\prime}=\iota(\Ts_{1})$, $\Pst_{1}^{\prime}=\iota(\Pst_{1})$\}
 (under the realization \eqref{tri-real})
 satisfy \eqref{tri-alg} with replacement $k_{+} \to k_{-}$ and are 
  right co-ideals  of $U_{q}(\widehat{sl_{2}})$ with respect to the co-product \eqref{copro-h}.
\subsection{Solutions of the intertwining relations}
The intertwining relations (equivalent to \eqref{int-0}) associated with the triangular q-Onsager algebra are given by:
\begin{align}
\mathsf{ev}_{x^{-1}}(a)\Kf(x) &= 
 \Kf(x)\mathsf{ev}_{x}(a) 
 \quad \text{for} \quad a \in \{\Ts_{0},\Ts_{1}, \Pst_{1} \} ,
 \label{int-1}
\end{align}
where we assume that the spectral parameter is of the form $x=q^{u}$, $u \in {\mathbb C}$. 
We assume that 
the K-operator $ \Kf(x)$ is a power series on generators of $U_{q}(sl_{2})$. 
%
%
Explicitly, we have 
\begin{align}
 ( k_{+}qx^{-s_1} Eq^{H}   + \epsilon_{+} q^{H})
 \Kf(x)
&=
\Kf(x)
 ( k_{+}qx^{s_1} Eq^{H}   + \epsilon_{+} q^{H}  )  , 
 \label{int3}
 \\[6pt]
 (  k_{+}x^{s_0}E + \epsilon_{-} q^{-H} ) 
 \Kf(x)
&=
\Kf(x)
 (   k_{+}x^{-s_0}E + \epsilon_{-} q^{-H}  ) ,
 \label{int4}
\end{align}
and 
\begin{multline}
\left(-(q^{2}-q^{-2})F(\epsilon_{-}qx^{s_{1}}q^{-H}+\
  \epsilon_{+}x^{-s_{0}}) +
  k_{+}\frac{q+q^{-1}}{q-q^{-1}}(x^{-s}q^{H}+x^{s}q^{-H}) 
  \right)\Kf(x)=
\\
=\Kf(x)\left(-(q^{2}-q^{-2})F(\epsilon_{-}qx^{-s_{1}}q^{-H}+\
  \epsilon_{+}x^{s_{0}}) +
  k_{+}\frac{q+q^{-1}}{q-q^{-1}}(x^{s}q^{H}+x^{-s}q^{-H}) 
  \right).
  \label{int5}
\end{multline}
In the third relation \eqref{int5}, we omit the trivial contribution from the terms on the central element and the scalar. 
We will prove the following theorem.
\begin{theorem}
The following K-operator solves the intertwining relations \eqref{int-1}.
\footnote{The formula for $|q|<1$ can be obtained by replacement: 
$\exp_{q^{-2}}(x)=((1-q^{-2})x;q^{-2})^{-1}_{\infty}\to \exp^{-1}_{q^{2}}(-x)=((q^{2}-1)x;q^{2})_{\infty}$. 
We also remark that \eqref{Kgen-up} can be rewritten as $\Kf(x)={\mathfrak K}(x^{-1})^{-1}x^{s_{0}H}{\mathfrak K}(x)$, where 
${\mathfrak K}(x)=\exp_{q^{-2}}\left( -\frac{\epsilon_{-}}{\epsilon_{+}(q-q^{-1})}x^{-s}q^{-H} \right)
\exp_{q^{-2}}\left( -\frac{qk_{+}}{\epsilon_{-}(q-q^{-1})}x^{-s_{0}}Eq^{H} \right)  $.}
\begin{align}
 \Kf(x)&=x^{s_{0}H}
 \exp_{q^{-2}}^{-1}\left( \alpha_{+} Eq^{H} \right)
 \frac{\left( -\frac{\epsilon_{-}}{\epsilon_{+}}x^{s}q^{-H-1};q^{-2} \right)_{\infty}}{\left( -\frac{\epsilon_{-}}{\epsilon_{+}}x^{-s}q^{-H-1};q^{-2} \right)_{\infty}}
  \exp_{q^{-2}}\left(  \alpha_{+}Eq^{H} \right)
\nonumber \\[6pt]
&= x^{s_{0}H}
 \frac{\left( -\frac{q^{-1}x^{s}}{\epsilon_{+}}(k_{+}x^{-s_{0}}E+\epsilon_{-}q^{-H});q^{-2} \right)_{\infty}}
 {\left( -\frac{q^{-1}x^{-s}}{\epsilon_{+}}(k_{+}x^{-s_{0}}E+\epsilon_{-}q^{-H});q^{-2} \right)_{\infty}}
  \qquad 
  \text{for} \quad k_{-}=0, \quad |q|>1,
  \label{Kgen-up}
 \end{align}
 where we set $\alpha_{+}=-\frac{qk_{+}x^{-s_{0}}}{(q-q^{-1})\epsilon_{-}}$, 
  $x=q^{u}$, $x^{s_{0}H}=q^{us_{0}H}$, $u \in {\mathbb C}$.
 \end{theorem}
{\em Proof.} 
First we rewrite \eqref{int3} and \eqref{int4}  in terms of 
$\Kf(x)=x^{s_{0}H}\widetilde{\Kf}(x)$ to get 
\begin{align}
 ( k_{+}qx^{-s_0-s} Eq^{H}  +  \epsilon_{+} q^{H})
 \widetilde{\Kf}(x)
&=
\widetilde{\Kf}(x)
 ( k_{+}qx^{s_1} Eq^{H}  +  \epsilon_{+} q^{H}  )  , 
 \label{int3p}
 \\[6pt]
 ( k_{+}x^{-s_0}E + \epsilon_{-} q^{-H} ) 
\widetilde{\Kf}(x)
&=
\widetilde{\Kf}(x)
 (   k_{+}x^{-s_0}E + \epsilon_{-} q^{-H}  ) . 
 \label{int4p}
\end{align}
We find that $\mathsf{ev}_{x}(\Ts_{1})$ is  transformed to a Cartan element of $U_{q}(sl_{2})$ by the following similarity 
transformation (cf. \eqref{E-h-Ei}, \eqref{E-E-Ei})
\footnote{A similar procedure was used in \cite{FGK19} for the rational case.}: 
\begin{align}
\exp_{q^{-2}}\left( \alpha_{+} Eq^{H} \right)
(   k_{+}x^{-s_0}E + \epsilon_{-} q^{-H}  )
\exp_{q^{-2}}^{-1}\left( \alpha_{+} Eq^{H} \right)
=\epsilon_{-} q^{-H} ,
\label{simE}
\end{align}
  where $\alpha_{+}=-\frac{qk_{+}x^{-s_{0}}}{(q-q^{-1})\epsilon_{-}}$. 
 Thus 
$\widetilde{\Kf}(x)=\exp_{q^{-2}}^{-1}\left( \alpha_{+} Eq^{H} \right)\Kf_{0}(x)\exp_{q^{-2}}\left( \alpha_{+} Eq^{H} \right)$ 
solves \eqref{int4p} if  
$\Kf_{0}(x)$ commutes with any Cartan element of $U_{q}(sl_{2})$. 
In this case, 
\eqref{int3p} boils down to 
\begin{align}
E(\epsilon_{+}q^{1+H}+\epsilon_{-}x^{-s})\Kf_{0}(x)=
\Kf_{0}(x)E(\epsilon_{+}q^{1+H}+\epsilon_{-}x^{s}), 
\label{intAE}
\end{align}
which is equivalent to an intertwining relation for 
 generic diagonal solutions of the reflection equation (eq.\ (4.8) in \cite{BT18}).  
 Making use of a generic diagonal solution
\footnote{There is a freedom on an overall factor $f(q^{\xi H})$ for a fixed $\xi \in {\mathbb C}$
which commutes with the generator $E$.  Here $f(y)$ 
 is a series on $y \in {\mathbb C}$ whose coefficients are 
central elements of $U_{q}(sl_{2})$ or complex numbers and satisfies $f(yq^{2\xi})=f(y)$. 
}
\begin{align}
\Kf_{0}(x)&= \frac{\left( -\frac{\epsilon_{-}}{\epsilon_{+}}x^{s}q^{-H-1};q^{-2} \right)_{\infty}}{\left( -\frac{\epsilon_{-}}{\epsilon_{+}}x^{-s}q^{-H-1};q^{-2} \right)_{\infty}}
\nonumber \\[6pt]
&= \exp_{q^{-2}}^{-1}\left( -\frac{\epsilon_{-}}{\epsilon_{+}(q-q^{-1})}x^{s}q^{-H} \right)
\exp_{q^{-2}}\left( -\frac{\epsilon_{-}}{\epsilon_{+}(q-q^{-1})}x^{-s}q^{-H} \right) 
\quad \text{for} 
\quad |q|>1, 
 \label{Kdia}
\end{align}
which satisfies \eqref{intAE} 
 ($x^{s_{0}H}\Kf_{0}(x)$ corresponds to eq.\ (4.9) in \cite{BT18}), 
 we find that \eqref{Kgen-up} satisfies \eqref{int3} and \eqref{int4}. 

 Next, we show that \eqref{Kgen-up} satisfies \eqref{int5}. 
 Making use of \eqref{E-h-Ei} and \eqref{E-F-Ei},  and taking note on the fact that $\Kf_{0}(x)$ 
 commutes with the Cartan element and the central element, 
 we find that \eqref{int5} reduces to a combination of \eqref{intAE} and 
 \begin{align}
F(\epsilon_{+}+\epsilon_{-}x^{s}q^{1-H})\Kf_{0}(x)=
\Kf_{0}(x)F(\epsilon_{+}+\epsilon_{-}x^{-s}q^{1-H}) ,
\label{intAF}
\end{align}
 which is equivalent to another intertwining relation for 
 generic diagonal solutions of the reflection equation (eq.\ (4.7) in \cite{BT18}; 
 thus \eqref{Kdia} satisfies \eqref{intAF}.) 
 The second equality of \eqref{Kgen-up} follows from 
 the inverse of \eqref{simE}. \hspace{\fill} $\square$

 The solution \eqref{Kgen-up} can be well defined 
   at least for finite dimensional representations of $U_{q}(sl_{2})$ 
 since there is a finite $n$  such that $(Eq^{H})^{n}=0$, 
 and $q^{-H}$ and $x^{s_{0}H}$ are finite size diagonal matrices (on  
 appropriate base vectors).  
 In particular for the fundamental representation,  the solution 
 \eqref{Kgen-up} reproduces the $2 \times 2$ triangular matrix solution 
 (\eqref{Kmat-sc} for $k_{-}=0$) of the reflection equation 
 \eqref{refeq0}:
\begin{align}
\pi(\Kf(x))&=
\kappa(x)
\begin{pmatrix}
x^{s_{0}}\epsilon_{+} + x^{-s_{1}} \epsilon_{-} & 
\frac{k_{+}(x^{s}-s^{-s})}{q-q^{-1}} \\
0  & 
x^{-s_{0}}\epsilon_{+} + x^{s_{1}} \epsilon_{-}
 \end{pmatrix} ,
 \label{K-mat1}
 \end{align}
 where $ \kappa(x)$ is an overall factor defined by 
 \begin{align}
  \kappa(x)=
 \frac{
      \left(-\frac{\epsilon_{-}}{\epsilon_{+}}x^{s}q^{-2};q^{-2}
   \right)_{\infty}
   }{
       \epsilon_{+} \left(-\frac{\epsilon_{-}}{\epsilon_{+}}x^{-s};q^{-2}
   \right)_{\infty}
   }.  
   \label{overallK}
 \end{align}
  \subsection{Solutions of the reflection equation}
We remark that the intertwining relations 
$\mathbf{r}_{i}((\mathsf{ev}_{x} \otimes \mathsf{ev}_{y})\Delta^{\prime}(a))=
((\mathsf{ev}_{x^{-1}} \otimes \mathsf{ev}_{y^{-1}})\Delta^{\prime}(a)) \mathbf{r}_{i}$, $i=1,2$,  
for any $a \in O^{t}_{q}(\widehat{sl_{2}})$
follow 
from \eqref{intert1}, \eqref{intert2} and \eqref{int-1}, 
where the right hand side and the left hand side of 
\eqref{refeq2} are denoted
\footnote{Here $\Kf(x)$ in $\mathbf{r}_{i}$ is assumed to be the one  in \eqref{int-1}.}
 as 
$\mathbf{r}_{1} $ and $\mathbf{r}_{2} $, respectively. 
This suggests the following theorem.
\begin{theorem}
The K-operator 
\eqref{Kgen-up} is a solution of the 
reflection equation \eqref{refeq2}. 
\end{theorem}
We have proven this by lengthy direct computation. 
\\
{\em Proof.} 
Expanding \eqref{refeq2} with respect to $y$ and multiplying Cartan elements on both sides, we find four different types of relations on 
$\Kf(x)$, two of which are identical to \eqref{int3} and \eqref{int4}, and the other two have the form:
\begin{multline}
-(q-q^{-1})^{2}x^{-s_{1}}
(k_{+}q^{-1}x^{s_{0}}E+\epsilon_{-}q^{-1-H}+\epsilon_{+}x^{s})\Kf(x)Fq^{-H}
\\
+(q-q^{-1})^{2}x^{s_{1}}Fq^{-H}\Kf(x)
(k_{+}qx^{-s_{0}}E+\epsilon_{-}q^{1-H}+\epsilon_{+}x^{-s})
\\
+k_{+}(x^{s}-x^{-s}q^{-2})(\Kf(x)-q^{-H}\Kf(x)q^{-H})=0
\label{int-re1}
\end{multline}
and 
\begin{align}
Eq^{H}\Kf(x)(\epsilon_{+}x^{s_{0}}q^{H+1}+\epsilon_{-}x^{-s_{1}})=
(\epsilon_{+}x^{-s_{0}}q^{H-1}+\epsilon_{-}x^{s_{1}})\Kf(x)Eq^{H} .
\label{int-re2}
\end{align}
We show that the assumption  \eqref{Kgen-up} satisfies \eqref{int-re1} and \eqref{int-re2} leads to trivial identities. 
First, we apply \eqref{int4} for \eqref{int-re1} and replace $FE$ and $EF$ with  Cartan elements and the Casimir element 
based on \eqref{Casimir}. Then we take the term containing $F$ to the right side of $\Kf(x)$ by \eqref{int5}, and 
derive an equation on $\Kf_{0}(x)$ by \eqref{E-h-Ei} and \eqref{E-F-Ei}. Taking note on the relation
\begin{align}
E \Kf_{0}(x)=\Kf_{0}(x) 
\left(\frac{1+\frac{\epsilon_{-}}{\epsilon_{+}}x^{s}q^{1-H}}{1+\frac{\epsilon_{-}}{\epsilon_{+}}x^{-s}q^{1-H}} \right) E,
\label{H-K}
\end{align}
which follows from \eqref{Kdia}, we arrive at the following relation
\begin{multline}
\Kf_{0}(x)\Bigl[ -
\left(q^H-\lambda \alpha_{+}\, \frac{\epsilon_{+}+\epsilon_{-}x^{s}q^{1-H}}{\epsilon_{+}+\epsilon_{-}x^{-s}q^{1-H}}\, Eq^{2H+1}  \right)
\Bigl\{
\left(\lambda k_{+}q^{-1}x^{-s}- \lambda^{2}\epsilon_{+}x^{s_{0}}\alpha_{+}q^{H-1}\right)C
+  \\
\lambda F \left(\epsilon_{+}x^{s_{0}}+ \epsilon_{-}x^{-s_{1}}q^{1-H} \right)
-\alpha_{+} E \left(\lambda \epsilon_{+}\alpha_{+}x^{s_{0}}q^{3H+1}- k_{+}x^{-s}q^{2H-1} \right)
+\epsilon_{+}\alpha_{+}(1+q^{2})x^{s_{0}}q^{2H-2}
\\
+\epsilon_{-}\alpha_{+}x^{-s_{1}}q^{H-1}-k_{+}\lambda^{-1}x^{-s}q^{H-2}
\Bigr\}
+\Bigl\{
\left(\lambda k_{+}q^{-1}x^{s}- \lambda^{2}\epsilon_{+}x^{s_{0}}\alpha_{+}q^{H-3}\right)C+
\\
\lambda q^{-2}F \left(\epsilon_{+}x^{s_{0}}+ \epsilon_{-}x^{-s_{1}}q^{1-H} \right)
-\alpha_{+} E \left(\lambda \epsilon_{+}\alpha_{+}x^{s_{0}}q^{3H-1}- k_{+}x^{s}q^{2H-1} \right)
+\epsilon_{+}\alpha_{+}(1+q^{2})x^{s_{0}}q^{2H-4}
\\
+\epsilon_{-}\alpha_{+}x^{-s_{1}}q^{H-3}-k_{+}\lambda^{-1}x^{s}q^{H-2}
\Bigr\} \left(q^{H}+\alpha_{+}(1-q^{2})Eq^{2H}\right)
-\lambda^{-1}k_{+}\left(x^{s}-x^{-s}q^{-2}\right)
\Bigr] =0, 
\\
\lambda=q-q^{-1}.
\label{K0=0}
\end{multline} 
Expansion of 
the part $[\dots ]$  in \eqref{K0=0}
 is lengthy and involved; but one can check that it is indeed $0$ if 
 one replaces $FE$ and $EF$ with  Cartan elements and the Casimir element 
based on \eqref{Casimir}. 
By using \eqref{E-h-Ei}, one can show that \eqref{int-re2} reduces to \eqref{intAE} 
(or  one may use \eqref{H-K}). \hspace{\fill} $\square$ 

 The solution of the reflection equation \eqref{refeq2} associated with another triangular solution \eqref{Kmat-sc} with $k_{+}=0$ 
 follows from \eqref{Kgen-up}  
 \begin{align}
 \Kf(x)&=\iota(\eqref{Kgen-up})=x^{s_{0}H}
 \exp_{q^{-2}}\left( \alpha_{-} F\right)
 \frac{\left( -\frac{\epsilon_{-}}{\epsilon_{+}}x^{s}q^{-H-1};q^{-2} \right)_{\infty}}{\left( -\frac{\epsilon_{-}}{\epsilon_{+}}x^{-s}q^{-H-1};q^{-2} \right)_{\infty}}
  \exp_{q^{-2}}^{-1} \left(  \alpha_{-}F \right)
\nonumber \\[6pt]
&= x^{s_{0}H}
 \frac{\left( -\frac{q^{-1}x^{s}}{\epsilon_{+}}(k_{-}qx^{s_{0}}Fq^{-H}+\epsilon_{-}q^{-H});q^{-2} \right)_{\infty}}
 {\left( -\frac{q^{-1}x^{-s}}{\epsilon_{+}}(k_{-}qx^{s_{0}}Fq^{-H}+\epsilon_{-}q^{-H});q^{-2} \right)_{\infty}}
\quad  
  \text{for} \quad k_{+}=0,  \quad |q|>1,
  \label{Kgen-lo}
 \end{align}
 where $\alpha_{-}=q^{-1}x^{2s_{0}}\iota(\alpha_{+})=-\frac{k_{-}x^{s_{0}}}{(q-q^{-1})\epsilon_{-}}$.
 In particular for the fundamental representation,  the solution 
 \eqref{Kgen-lo} reproduces the $2 \times 2$ triangular matrix solution 
 (\eqref{Kmat-sc} for $k_{+}=0$) of the refection equation 
 \eqref{refeq0}:
\begin{align}
\pi (\Kf(x))&=
\kappa(x)
\begin{pmatrix}
x^{s_{0}}\epsilon_{+} + x^{-s_{1}} \epsilon_{-} & 
0 \\
\frac{k_{-}(x^{s}-s^{-s})}{q-q^{-1}}  & 
x^{-s_{0}}\epsilon_{+} + x^{s_{1}} \epsilon_{-}
 \end{pmatrix} ,
 \label{K-mat2}
 \end{align}
 where $ \kappa(x)$ is the overall factor defined by \eqref{overallK}.
 
 We obtain two more solutions of  the reflection equation \eqref{refeq2} 
 with a different prefactor. The first one is 
 \begin{align}
 \Kf(x)&=\sigma(\eqref{Kgen-up})=x^{-s_{1}H}
 \exp_{q^{-2}}^{-1}\left( \beta_{-} Fq^{-H} \right)
 \frac{\left( -\frac{\epsilon_{+}}{\epsilon_{-}}x^{s}q^{H-1};q^{-2} \right)_{\infty}}{\left( -\frac{\epsilon_{+}}{\epsilon_{-}}x^{-s}q^{H-1};q^{-2} \right)_{\infty}}
  \exp_{q^{-2}}\left(  \beta_{-}Fq^{-H} \right)
\nonumber \\[6pt]
&= x^{-s_{1}H}
 \frac{\left( -\frac{q^{-1}x^{s}}{\epsilon_{-}}(k_{-}x^{-s_{1}}F+\epsilon_{+}q^{H});q^{-2} \right)_{\infty}}
 {\left( -\frac{q^{-1}x^{-s}}{\epsilon_{-}}(k_{-}x^{-s_{1}}F+\epsilon_{+}q^{H});q^{-2} \right)_{\infty}}
\quad   \text{for} \quad k_{+}=0, \quad |q|>1, 
  \label{Kgen-upd}
 \end{align}
 where $\beta_{-}=\sigma(\alpha_{+})=-\frac{qk_{-}x^{-s_{1}}}{(q-q^{-1})\epsilon_{+}}$. 
The second one is 
\begin{align}
 \Kf(x)&=\sigma(\iota(\eqref{Kgen-up}))=\sigma(\eqref{Kgen-lo})
 \nonumber 
 \\[6pt]
 &=x^{-s_{1}H}
 \exp_{q^{-2}}\left( \beta_{+} E\right)
 \frac{\left( -\frac{\epsilon_{+}}{\epsilon_{-}}x^{s}q^{H-1};q^{-2} \right)_{\infty}}{\left( -\frac{\epsilon_{+}}{\epsilon_{-}}x^{-s}q^{H-1};q^{-2} \right)_{\infty}}
  \exp_{q^{-2}}^{-1} \left(  \beta_{+}E \right)
\nonumber \\[6pt]
&= x^{-s_{1}H}
 \frac{\left( -\frac{q^{-1}x^{s}}{\epsilon_{-}}(k_{+}qx^{s_{1}}Eq^{H}+\epsilon_{+}q^{H});q^{-2} \right)_{\infty}}
 {\left( -\frac{q^{-1}x^{-s}}{\epsilon_{-}}(k_{+}qx^{s_{1}}Eq^{H}+\epsilon_{+}q^{H});q^{-2} \right)_{\infty}}
   \qquad 
  \text{for} \quad k_{-}=0,  \quad |q|>1,
  \label{Kgen-lou}
 \end{align}
 where $\beta_{+}=\sigma(\alpha_{-})=-\frac{k_{+}x^{s_{1}}}{(q-q^{-1})\epsilon_{+}}$.

We remark that our generic triangular solutions \eqref{Kgen-up}, \eqref{Kgen-lo}, \eqref{Kgen-upd} and \eqref{Kgen-lou} 
reduce to the generic diagonal solutions \cite{BT18} at $k_{+}=k_{-}=0$.
 \section{Discussion}
 In this paper, we solved the intertwining relations of the triangular q-Onsager algebra 
under the evaluation map, and obtained generic triangular solutions of the reflection 
equation. A natural problem in this direction will be to generalize our solutions 
to the most general case $k_{+}k_{-} \ne 0$, which is related to the q-Onsager algebra \cite{Te03,Ba04}.  
 The q-Onsager algebra $O_q(\widehat{sl_2})$   (with the central element $\Gamma=1$)   
 is generated by two elements ${\textsf W}_0,{\textsf W}_1$ and  unit obeying the following relations.
\begin{align}
\begin{split}
 \big[ \Ws_0, \big[  \Ws_0, \big[ \Ws_0, \Ws_1 \big]_{q^{-2}} \big]_{q^{2}} \big]
 & = (q+q^{-1})^{2} k_{+} k_{-} \big[ \Ws_0, \Ws_1 \big] , 
 \\[6pt]
\big[ \Ws_1,  \big[ \Ws_1, \big[ \Ws_1, \Ws_0 \big]_{q^{-2}} \big]_{q^{2}} \big]
& = (q+q^{-1})^{2} k_{+} k_{-} \big[ \Ws_1, \Ws_0 \big] .
\end{split}
\end{align}
This algebra is realized
\footnote{
The convention used in eq.\ (2.7)  in the paper \cite{BB17} is related to \eqref{al-W} 
by the automorphism  \eqref{auto2} of $U_{q}(\widehat{sl_{2}})$ and
the replacement $k_{\pm}\to k_{\mp}$ and $q \to q^{-1}$
(under the condition $h_{0}+h_{1}=c=0$, $\Gamma=1$.)}
 in terms of the generators of $U_{q}(\widehat{sl_{2}})$. 
\begin{align}
\begin{split}
\Ws_0&= k_{+}qe_{1}q^{h_{1}} + k_{-}f_1 + \epsilon_{+} q^{h_1},
 \\[6pt]
\Ws_1&= k_{-}qe_{0}q^{h_{0}} + k_{+}f_0 + \epsilon_{-}q^{h_0}.
\end{split}
\label{al-W}
\end{align}
Note that two of the generators \eqref{tri-real} of the triangular q-Onsager algebra are 
specializations of these: $\Ts_{0}=\Ws_{0}|_{k_{-}=0}$, $\Ts_{1}=\Ws_{1}|_{k_{-}=0}$. 
In order to obtain generic non-diagonal solutions of the reflection equation \eqref{refeq2} 
for  $k_{+}k_{-}\ne 0$, we must solve the intertwining relations
\begin{align}
\mathsf{ev}_{x^{-1}}(\Ws_{0})\Kf(x)&=\Kf(x) \mathsf{ev}_{x}(\Ws_{0}) ,
\label{int-O0}
\\[6pt]
\mathsf{ev}_{x^{-1}}(\Ws_{1})\Kf(x)&=\Kf(x) \mathsf{ev}_{x}(\Ws_{1}) .
\label{int-O1}
\end{align}
Our generic triangular solutions are deformation of the generic diagonal solution 
($k_{+}=k_{-}=0$ case)
\footnote{The generic diagonal solution corresponds to $x^{s_{0}H}\Kf_{0}(x)$, where $\Kf_{0}(x)$ is defined in \eqref{Kdia}.}
 \cite{BT18}. In view of this, 
it would be tempting to see if the following substitution works.
  \begin{align}
 \Kf(x)&= 
 x^{s_{0}H}
 \frac{\left( -\frac{q^{-1}x^{s}}{\epsilon_{+}} \mathsf{ev}_{x}(\Ws_{1});q^{-2} \right)_{\infty}}
 {\left( -\frac{q^{-1}x^{-s}}{\epsilon_{+}}\mathsf{ev}_{x}(\Ws_{1}); q^{-2} \right)_{\infty}}
 \nonumber 
\\[6pt]
& =
 x^{s_{0}H}
 \frac{\left( -\frac{q^{-1}x^{s}}{\epsilon_{+}}(k_{+}x^{-s_{0}}E+k_{-}qx^{s_{0}}Fq^{-H}+\epsilon_{-}q^{-H});q^{-2} \right)_{\infty}}
 {\left( -\frac{q^{-1}x^{-s}}{\epsilon_{+}}(k_{+}x^{-s_{0}}E+k_{-}qx^{s_{0}}Fq^{-H}+\epsilon_{-}q^{-H});q^{-2} \right)_{\infty}}
 \quad 
  \text{for}  \quad |q|>1.
  \label{K-On}
 \end{align}
In fact, this reduces to \eqref{Kgen-up} at $k_{-}=0$ and \eqref{Kgen-lo} at $k_{+}=0$. 
However,  \eqref{K-On} seems not satisfy \eqref{int-O0}, although 
 it solves \eqref{int-O1}. How this could be resolved  remains to be clarified. 
 We expect that 
 recent solutions of the reflection equation for the symmetric tensor 
 representations of $U_{q}(A^{(1)}_{n-1})$ \cite{KOY18}
 \footnote{See also \cite{RV18}, in which
  all invertible solutions of the reflection equation are classified for the vector representation, including all triangular ones.}
  (see also, \cite{ML19}) 
 and the rational case $Y(sl_{2})$ \cite{FGK19} are great clues for this. 
\section*{Acknowledgments} 
The author would like to thank Pascal Baseilhac for discussions. 
He also thanks the anonymous referees for their useful comments. 
  The work of the author was 
 supported in part by   
Osaka City University Advanced Mathematical Institute (MEXT Joint Usage/Research Center on  Mathematics and Theoretical Physics). 

\section*{Appendix:  q-analogue of Hadamard formula}
\label{appA}
\addcontentsline{toc}{section}{Appendix}
\def\theequation{A\arabic{equation}}
\setcounter{equation}{0}
In this section, we review useful formulas derived from the q-deformed Hadamard formula
\cite{KT91} for elements ${\mathbf A}$ and ${\mathbf B}$ in $U_{q}(sl_{2})$: 
\begin{align}
\exp_{q} ({\mathbf A}) {\mathbf B} \exp_{q}^{-1} ({\mathbf A})=
\sum_{k=0}^{\infty} \frac{{\mathbf B}_{k}}{(k)_{q} !},
\end{align}
where  
${\mathbf B}_{0}={\mathbf B}$, ${\mathbf B}_{k+1}=[{\mathbf A},{\mathbf B}_{k}]_{q^{k}}$. 
For $a,b,c \in {\mathbb C}$, 
one can prove the following relations (as formal series) based on induction and the property of the q-exponential funciton 
$\exp_{q}^{-1}(x)=\exp_{q^{-1}}(-x)$.
\begin{align}
 &\exp_{q^{-2}}\left( a Eq^{bH} \right)
 q^{cH}
  \exp_{q^{-2}}^{-1}\left(  a Eq^{bH} \right)
 =\sum_{j=0}^{\infty} \frac{(q^{2c};q^{-2})_{j}}{(q^{-2};q^{-2})_{j}}
 (a(1-q^{-2})Eq^{bH})^{j}q^{cH},
 \label{E-h-Ei}
\\[6pt]
& \exp_{q^{-2}}\left( a Eq^{bH} \right)
 Eq^{cH}
  \exp_{q^{-2}}^{-1}\left(  a Eq^{bH} \right)
 =\sum_{j=0}^{\infty} \frac{(q^{2(c-b)};q^{-2})_{j}}{(q^{-2};q^{-2})_{j}}
 (a(1-q^{-2})Eq^{bH})^{j}Eq^{cH}, 
 \label{E-E-Ei}
\\[6pt]
& \exp_{q^{-2}}\left( a Eq^{bH} \right)
 Fq^{cH}
  \exp_{q^{-2}}^{-1}\left(  a Eq^{bH} \right)
 =Fq^{cH}+
a(1-q^{-2})q^{-2b}\sum_{j=1}^{\infty} 
\frac{(a(1-q^{-2})Eq^{bH})^{j-1}}{(q^{-2};q^{-2})_{j}}
 \nonumber  \\
 & \ \times  
 \left\{
 (q^{2(b+c)};q^{-2})_{j}Cq^{(b+c)H} - 
 \frac{(q^{2(b+c+1)};q^{-2})_{j}q^{(b+c+1)H-1}+(q^{2(b+c-1)};q^{-2})_{j}q^{(b+c-1)H+1}}{(q-q^{-1})^{2}}
 \right\},
\label{E-F-Ei}
 \\[6pt]
 &\exp_{q^{-2}}^{-1}\left( a Eq^{bH} \right)
 q^{cH}
  \exp_{q^{-2}}\left(  a Eq^{bH} \right)
 =\sum_{j=0}^{\infty} \frac{(q^{2c};q^{2})_{j}}{(q^{2};q^{2})_{j}}
 (-a(1-q^{2})Eq^{bH})^{j}q^{cH},
 \label{Ei-h-E}
\\[6pt]
& \exp_{q^{-2}}^{-1}\left( a Eq^{bH} \right)
 Eq^{cH}
  \exp_{q^{-2}}\left(  a Eq^{bH} \right)
 =\sum_{j=0}^{\infty} \frac{(q^{2(c-b)};q^{2})_{j}}{(q^{2};q^{2})_{j}}
 (-a(1-q^{2})Eq^{bH})^{j}Eq^{cH}, 
 \label{Ei-E-E}
\\[6pt]
& \exp_{q^{-2}}^{-1}\left( a Eq^{bH} \right)
 Fq^{cH}
  \exp_{q^{-2}}\left(  a Eq^{bH} \right)
 =Fq^{cH}-
a(1-q^{2})q^{-2b}\sum_{j=1}^{\infty} 
\frac{(-a(1-q^{2})Eq^{bH})^{j-1}}{(q^{2};q^{2})_{j}}
 \nonumber  \\
 & \ \times  
 \left\{
 (q^{2(b+c)};q^{2})_{j}Cq^{(b+c)H} - 
 \frac{(q^{2(b+c-1)};q^{2})_{j}q^{(b+c-1)H+1}+(q^{2(b+c+1)};q^{2})_{j}q^{(b+c+1)H-1}}{(q-q^{-1})^{2}}
 \right\},
\label{Ei-F-E}
 \\[6pt]
& \exp_{q^{-2}}\left( a Fq^{bH} \right)
 q^{cH}
  \exp_{q^{-2}}^{-1}\left(  a Fq^{bH} \right)
 =\sum_{j=0}^{\infty} \frac{(q^{-2c};q^{-2})_{j}}{(q^{-2};q^{-2})_{j}}
 (a(1-q^{-2})Fq^{bH})^{j}q^{cH},
 \label{F-h-Fi}
\\[6pt]
& \exp_{q^{-2}}\left( a Fq^{bH} \right)
 Fq^{cH}
  \exp_{q^{-2}}^{-1}\left(  a Fq^{bH} \right)
 =\sum_{j=0}^{\infty} \frac{(q^{2(b-c)};q^{-2})_{j}}{(q^{-2};q^{-2})_{j}}
 (a(1-q^{-2})Fq^{bH})^{j}Fq^{cH},
 \label{F-F-Fi}
 \\[6pt]
& \exp_{q^{-2}} \left( a Fq^{bH} \right)
 Eq^{cH}
  \exp_{q^{-2}}^{-1}\left(  a Fq^{bH} \right) =
\nonumber \\
&\qquad  =Eq^{cH}+
a(1-q^{-2})q^{2b}\sum_{j=1}^{\infty} 
\frac{(a(1-q^{-2})Fq^{bH})^{j-1}}{(q^{-2};q^{-2})_{j}}  
 \Bigl\{
 (q^{-2(b+c)};q^{-2})_{j}Cq^{(b+c)H} 
  \nonumber  \\
 &
\qquad  \qquad 
 - \frac{(q^{-2(b+c-1)};q^{-2})_{j}q^{(b+c-1)H-1}+(q^{-2(b+c+1)};q^{-2})_{j}q^{(b+c+1)H+1}}{(q-q^{-1})^{2}}
 \Bigr\} ,
  \label{F-E-Fi}
 \\[6pt]
& \exp_{q^{-2}}^{-1}\left( a Fq^{bH} \right)
 q^{cH}
  \exp_{q^{-2}}\left(  a Fq^{bH} \right)
 =\sum_{j=0}^{\infty} \frac{(q^{-2c};q^2)_{j}}{(q^{2};q^{2})_{j}}
 (-a(1-q^{2})Fq^{bH})^{j}q^{cH},
 \label{Fi-h-F}
\\[6pt]
& \exp_{q^{-2}}^{-1}\left( a Fq^{bH} \right)
 Fq^{cH}
  \exp_{q^{-2}}\left(  a Fq^{bH} \right)
 =\sum_{j=0}^{\infty} \frac{(q^{2(b-c)};q^2)_{j}}{(q^2;q^2)_{j}}
 (-a(1-q^2)Fq^{bH})^{j}Fq^{cH} ,
 \label{Fi-F-F}
\\[6pt]
& \exp_{q^{-2}}^{-1} \left( a Fq^{bH} \right)
 Eq^{cH}
  \exp_{q^{-2}}\left(  a Fq^{bH} \right) =
\nonumber \\
&\qquad  =Eq^{cH}-
a(1-q^{2})q^{2b}\sum_{j=1}^{\infty} 
\frac{(-a(1-q^{2})Fq^{bH})^{j-1}}{(q^{2};q^{2})_{j}}
 \nonumber  \\
 & \ \times  
 \left\{
 (q^{-2(b+c)};q^{2})_{j}Cq^{(b+c)H} - 
 \frac{(q^{-2(b+c+1)};q^{2})_{j}q^{(b+c+1)H+1}+(q^{-2(b+c-1)};q^{2})_{j}q^{(b+c-1)H-1}}{(q-q^{-1})^{2}}
 \right\}.
 \label{Fi-E-F}
 \end{align} 
 The relation 
$(q^{a};q)_{k}=(-1)^{k} q^{ak+\frac{(k-1)k}{2}}(q^{-a};q^{-1})_{k}$ is useful to modify the above relations. 

\end{document}